\documentclass{nature_fig}

\usepackage{bbold}
\usepackage{bbm}
\usepackage[pdftex]{graphicx}
\usepackage{latexsym,amsmath,verbatim}
\usepackage{color}
\usepackage{rotating}
\usepackage{multirow}
\usepackage[english]{babel}
\usepackage{color}
\usepackage{changes}
\usepackage{todonotes}
\definechangesauthor[name={Manlio}, color=orange]{Manlio}

\definecolor{snorkelBlue}{rgb}{0,0.31,0.52}
\definecolor{peachColor}{rgb}{0.97,0.47,0.42}
\definecolor{Biscaybay}{rgb}{0, 0.556, 0.655}
\definecolor{lightGray}{rgb}{0.7,0.7,0.7}
\newcommand{\header}[1]{\noindent\textbf{\emph{#1.}}}
\newcommand{\rev}[1]{{\color{black} #1}}

\title{\rev{The physics of spreading processes in multilayer networks}}

\author{Manlio De Domenico$^{1\ast}$, Clara Granell$^{2}$, Mason A. Porter$^{3}$, and Alex Arenas$^{1\ast}$}

\begin{document}

\maketitle

\begin{affiliations}
\item Departament d'Enginyeria Inform\`{a}tica i Matem\`{a}tiques,
  Universitat Rovira i Virgili, 43007 Tarragona, Spain
\item Carolina Center for Interdisciplinary Applied Mathematics, Department of Mathematics, University of North Carolina, Chapel Hill, NC 27599-3250, USA  
\item Oxford Centre for Industrial and Applied Mathematics, Mathematical Institute, University of Oxford, OX2 6GG, UK; CABDyN Complexity Centre, University of Oxford, Oxford OX1 1HP, UK; and Department of Mathematics, University of California, Los Angeles, California 90095, USA
\end{affiliations}
\noindent Corresponding authors: alexandre.arenas@urv.cat, manlio.dedomenico@urv.cat
\date{}

%%%%%%%

%%%%%%%%%

%%%%%%%%%%%%%%%%% END OF PREAMBLE %%%%%%%%%%%%%%%%

%\begin{document}
\baselineskip24pt
%\maketitle

%%%%%%%%%%%%%%%%%%%%%%%%%%%%%%%%%%%%%%%%%%%%%%%%%%%%%%%%%%%%
%%
%%                                                                              A B S T R A C T
%%
%%%%%%%%%%%%%%%%%%%%%%%%%%%%%%%%%%%%%%%%%%%%%%%%%%%%%%%%%%%%

\begin{abstract}

The study of networks plays a crucial role in investigating the structure, dynamics, and function of a wide variety of complex systems in myriad disciplines. Despite the success of traditional network analysis, standard networks provide a limited representation of \rev{complex} systems, which often \rev{include} different types of relationships (i.e., ``multiplexity'') among their constituent components and/or multiple interacting subsystems. Such structural complexity has a significant effect on both dynamics and function. Throwing away or aggregating available structural information can generate misleading results and \rev{be} a major obstacle towards attempts to understand \rev{complex systems}.  
The recent ``multilayer'' approach for modeling networked systems explicitly allows the incorporation of multiplexity and other features of realistic \rev{systems}. On one hand, it allows one to couple different structural relationships by encoding them in a convenient mathematical object. On the other hand, it also allows one to couple different dynamical processes on top of such interconnected structures. The resulting framework plays a crucial role in helping \rev{achieve} a thorough, accurate understanding of complex systems. The study of multilayer networks has also revealed new physical phenomena that \rev{remain} hidden when using \rev{ordinary graphs, the traditional network representation}.
Here we survey progress towards \rev{attaining} a deeper understanding of \rev{spreading} processes on multilayer networks, and we highlight some of the physical phenomena \rev{ related to spreading processes that emerge from multilayer structure.}
\end{abstract}

%%%%%%%%%%%%%%%%%%%%%%%%%%%%%%%%%%%%%%%%%%%%%%%%%%%%%%%%%%%%
%%
%%                                                                              I N T R O D U C T I O N 
%%
%%%%%%%%%%%%%%%%%%%%%%%%%%%%%%%%%%%%%%%%%%%%%%%%%%%%%%%%%%%%

\section*{Introduction} 

Networks provide a powerful representation of interaction patterns in complex systems\cite{Newman2010Book,boccaletti2006complex,faust}. The structure of social relations among individuals, interactions between proteins, food webs, and many other situations can be represented using networks. Until recently, the vast majority of \rev{studies focused on networks} that consist of a single type of entity, with different entities connected to each other via a single type of connection. Such networks are now called {\em single-layer} (or {\em monolayer}) networks. The idea of incorporating additional information --- such as multiple types of interactions, subsystems, and time-dependence --- has long been pointed out in various fields, \rev{such as sociology, anthropology, and engineering,} but an effective unified framework for the mathematical treatment of such multidimensional structures, which are usually called {\em multilayer networks}, \rev{was} developed only recently\cite{kivela2014multilayer,boccaletti2014structure}.

Multilayer networks can be used to model many complex systems. For example, relationships between humans include different \rev{types} of interactions --- such as relationships between family members, friends, and coworkers --- that constitute different {\em layers} of a social system. Different layers of connectivity also arise naturally in natural and human-made systems in transportation\cite{gallotti2014anatomy}, ecology\cite{pilosof2015}, neuroscience\cite{bullmore2012economy}, and numerous other areas. The potential of multilayer networks for representing complex systems more accurately than was previously possible has led to an explosion of work on the physics of multilayer networks.

\rev{A key question concerns} the implications of multilayer structures on the dynamics of complex systems, and \rev{several papers} about interdependent networks --- a special type of multilayer network --- revealed that such structures can change the qualitative behaviors in a significant way. \rev{For example, several studies have provided insights on percolation properties and catastrophic cascades of failures in multilayer networks\cite{buldyrev2010catastrophic,gao2012networks,Baxter12,Son12,Bianconi14,Goh14,Hackett16}. These findings helped highlight an important challenge: How does one account for multiple layers of connectivity in a consistent mathematical way?}
An explosion of recent papers has developed the field of multilayer networks into its modern form, and there is now a suitable mathematical framework\cite{dedomenico2013mathematical}, novel structural descriptors\cite{Sola2013Centrality,halu2013multiplex,battiston2014structural,dedomenico2015ranking,cozzo2015triadic}, \rev{and tools from fields (such as statistical physics \cite{Bianconi13,Menichetti14})} for studying these systems. Many studies have also started to highlight the importance of analyzing multilayer networks, \rev{instead of relying on their monolayer counterparts, to gain new insights about empirical systems (see, e.g., \cite{cardillo2013emergence,braun2015reconfiguration}).}

\rev{It has now been recognized} that the study of multilayer networks is fundamental for enhancing understanding of dynamical processes on \rev{networked systems. An important example are spreading processes,} such as flows \rev{(and congestion)} in transportation \rev{networks\cite{morris2012transport,sole2016PRL},} and information and disease spreading in social networks\cite{bauch2015,funk2015,granell2013interplay,sanz2014dynamics,lima2015disease}. For instance, when two spreading process are coupled in a multilayer network, the onset of one disease-spreading process \rev{can depend} on the onset of the other one, \rev{and in some scenarios there is} a curve of critical points in the phase diagram of the parameters that govern a system's spreading dynamics\cite{granell2013interplay}. Such a curve reveals the existence of two distinct \rev{regimes, such that the criticality of the two dynamics is interdependent in one regime but not in the other.}
Similarly, cooperative behavior can be enhanced by multilayer structures, providing a novel way for cooperation to survive in structured populations\cite{gomez2012evolution}. For additional examples, see various reviews and surveys\cite{kivela2014multilayer,boccaletti2014structure,goh-review,salehi2015spreading,wang2015evolutionary,gao2012networks,bauch2015} on multilayer networks and specific topics within them.

A multilayer framework allows a natural representation of coupled structures and coupled dynamical processes. \rev{In this article,} after \rev{we give} a brief overview on \rev{representing} multilayer networks, we will focus on \rev{spreading} processes in which multilayer analysis has revealed new physical behavior. Specifically, we will discuss two cases: (i) a single dynamical process, such as continuous \rev{or} discrete diffusion, running \rev{on top of a multilayer network}; and (ii) different dynamical processes, \rev{in which each one runs on top of a given layer, but they are coupled by a multilayer structure.}

%%%%%%%%%%%%%%%%%%%%%%%%%%%%%%%%%%%%%%%%%%%%%%%%%%%%%%%%%%%%
%%
%%                S T R U C T U R A L    R E P R E S E N T A T I O N    O F    M U L T I L A Y E R    N E T W O R K S 
%%
%%%%%%%%%%%%%%%%%%%%%%%%%%%%%%%%%%%%%%%%%%%%%%%%%%%%%%%%%%%%

\section*{Structural representation of multilayer networks}

One can represent a monolayer network mathematically by using an adjacency \rev{matrix, which encodes} information about (possibly directed and/or weighted) relationships among \rev{the} entities in a network. Because multilayer networks include multiple dimensions of connectivity, called {\em aspects}, that have to be considered simultaneously, their structure is much richer than that of ordinary networks. Possible aspects include different types of interactions or communication channels, different subsystems, different spatial locations, different points in time, and more. One \rev{can use} tensors to encode the connectivity of multilayer networks as (multi)linear-algebraic objects\cite{dedomenico2013mathematical,kivela2014multilayer}. Multilayer networks include three types of edges: intra-layer edges (connecting nodes within the same layer), inter-layer edges between replica nodes (i.e., copies of the same entity) in different layers, and inter-layer edges between nodes that represent distinct entities. \rev{Distinguishing disparate types of edges has deep consequences both mathematically and physically. Mathematically, this yields banded structures in multilinear-algebraic objects that depend on a systems physical constraints, and such structures impact features such as a network's spectral properties. These, in turn, have a significant impact on dynamical systems (e.g., of spreading processes or coupled oscillators) that are coupled through multilayer networks. Moreover, intra-layer edges and inter-layer edges encode relationships in fundamentally different ways, and they thereby represent different types of physical functionality. For example, in a metropolitan transportation system \cite{gallotti2014anatomy,gallotti2015}, intra-layer edges account for connections between the same type of node (e.g., between two different subway stations), whereas inter-layer edges connect different types of nodes (e.g., between a certain subway station and an associated bus station). In some cases, inter-layer edges and intra-layer edges may even be measured using different physical units. For instance, an intra-layer edge in a multilayer social network could represent a friendship between two individuals on Facebook, whereas an inter-layer edge in the same network could represent the transition probability of an individual switching from using Facebook to use Twitter.}

\rev{The rich variety of connections in a typical multilayer network can be mathematically represented by the components $m_{i\alpha}^{j \beta}$ of a 4th-order tensor $\mathbf{M}$, called multilayer adjacency tensor\cite{dedomenico2013mathematical}, encoding the relationship between any node $i$ in layer $\alpha$ and any node $j$ in layer $\beta$ in the system (\rev{where} $i,j \in \{1,2,\dots,N\}$ and $\alpha,\beta \in \{1,2,\dots,L\}$, $N$ denotes the number of nodes in the network and $L$ denotes the number of layers).}

Once the connectivity of the nodes and layers are encoded in a tensor, one can define novel measures to characterize the multilayer structure. However, this is a delicate process, as naively generalizing existing concepts from monolayer networks can lead to qualitatively incorrect or nonsensical results\cite{kivela2014multilayer}. \rev{An alternative way of generalizing concepts from monolayer networks to multilayer networks is to use sets of adjacency matrices rather than tensors. This alternative approach has the advantage of familiarity, and indeed it is also convenient to ``flatten'' adjacency tensors into matrices (called ``supra-adjacency matrices'') for computations\cite{kivela2014multilayer,boccaletti2014structure}. However, the compact representation of multilayer networks in terms of tensors allows greater abstraction, which has been very insightful, and it will facilitate further development of the mathematics of complex systems.}

Studies of structural properties of multilayer networks include descriptors to identify the most \rev{``central''} nodes according to \rev{various notions} of importance\cite{dedomenico2013mathematical,Sola2013Centrality,halu2013multiplex,battiston2014structural,dedomenico2015ranking} and quantify triadic relations such as clustering and transitivity\cite{dedomenico2013mathematical,battiston2014structural,cozzo2015triadic}. Significant advances have been achieved to reduce the structural complexity \rev{of multilayer networks}\cite{dedomenico2015structural}, to unveil mesoscale structures (e.g., communities of densely-connected nodes) \cite{mucha2010community,Gauvin14,dedomenico2015identifying,peixoto2015inferring}, and to quantify intra-layer and inter-layer correlations\cite{min2014network,reis2014avoiding,nicosia2015correlations} in empirical networked systems.

The structural properties \rev{of multilayer networks} depend crucially on how layers are coupled together to form a multilayer structure. Inter-layer edges provide the coupling and help encode structural and dynamical features of a system, and their presence (or absence) produces fascinating structural and dynamical effects. For example, in multimodal transportation systems, in which layers represent different transportation modes, the weight of inter-layer connections might encode an economic or temporal cost to switching between two modes\rev{\cite{dedomenico2014navigability,gallotti2014anatomy}}. In multilayer social networks, inter-layer connections allow models to tune, in a natural way, an individual's self-reinforcement in opinion dynamics\cite{gardenes2015competition}. Depending on the relative \rev{importances} of intra-layer and inter-layer connections, a multilayer network can act either as a system of independent entities, in which layers are structurally decoupled, or as a single-layer system, in which layers are indiscernible in practice. In some multilayer networks, one can even derive a sharp transition between these two regimes\cite{radicchi2013abrupt,radicchi2014driving}.

%%%%%%%%%%%%%%%%%%%%%%%%%%%%%%%%%%%%%%%%%%%%%%%%%%%%%%%%%%%%
%%
%%                                 D Y N A M I C S     O N     M U L T I P L E X   N E T W O R K S 
%%
%%%%%%%%%%%%%%%%%%%%%%%%%%%%%%%%%%%%%%%%%%%%%%%%%%%%%%%%%%%%

\section*{Single and \rev{coupled} dynamics on multilayer networks}

\begin{figure}
\centering
\includegraphics[width=0.7\textwidth]{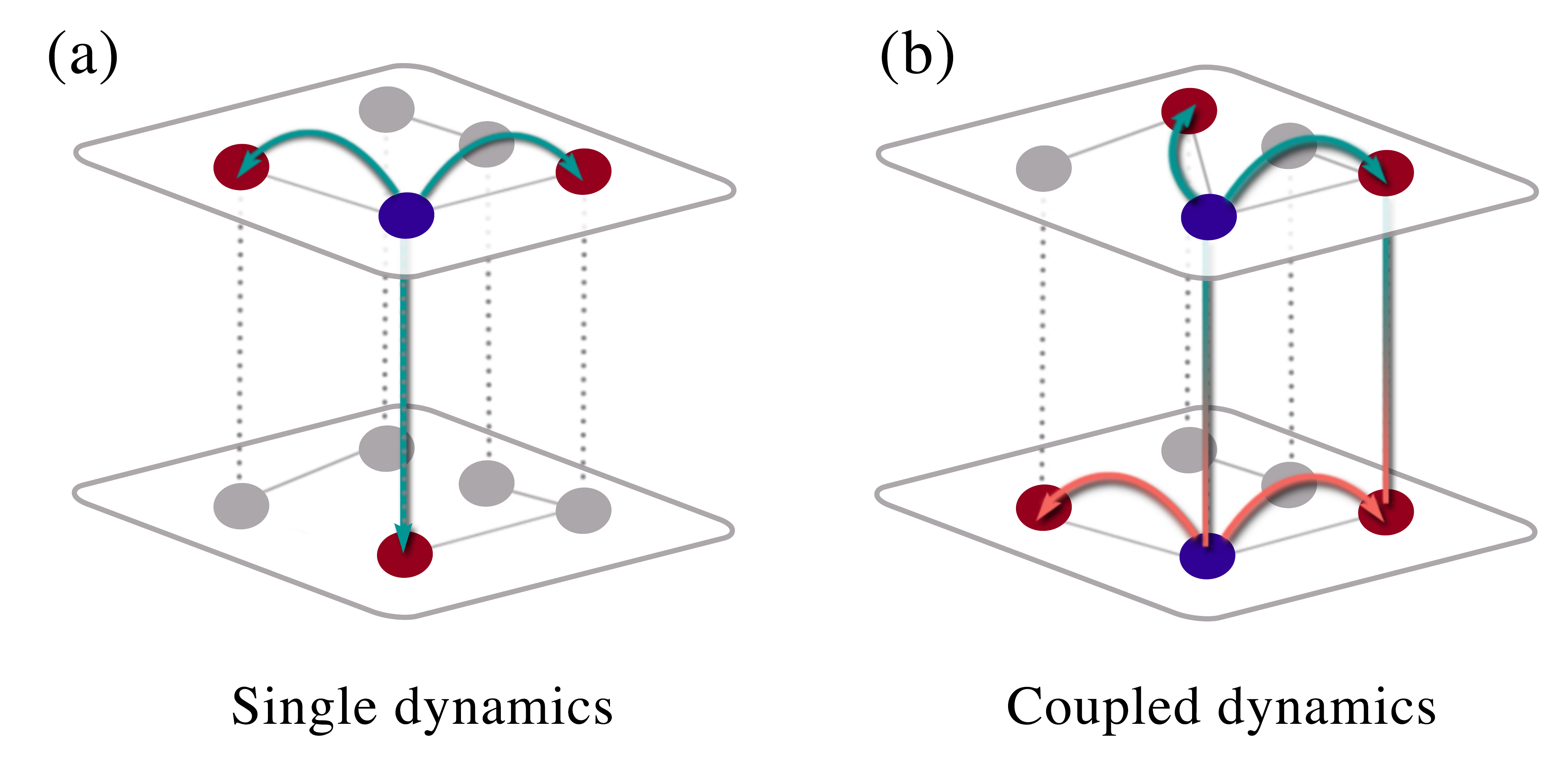}
\caption{\label{fig:dynamics_types}{\bf Dynamical processes on multilayer networks}. ({\bf a}) Schematic of a single type of dynamical process running on all layers of a multiplex network. (Arcs of the same color represent the same dynamical process.) ({\bf b}) Schematic of two dynamical processes, each of which is running on a different layer, that are coupled by the interconnected structure of a multilayer network. }
\end{figure}

There are two different categories of dynamical processes on multilayer networks: (i) a single dynamical process on top of the coupled structure of a multilayer network (see Fig.~\ref{fig:dynamics_types}a); and (ii) \rev{``mixed'' or ``coupled'' dynamics, in which} two or more dynamical processes are defined on each layer separately and are coupled together by the presence of inter-layer connections between nodes \rev{(see Fig.~\ref{fig:dynamics_types}b).}

\header{Single dynamics} In this section, we analyze physical phenomena that arise from a single dynamical process on top of a multilayer structure. The behavior of such a process depends both on intra-layer structure (i.e., the usual considerations in networks) and on inter-layer structure (i.e., the presence and strength of interactions between nodes on different layers). 

One of the simplest types of dynamics is a diffusion process (either continuous or discrete). The physics of diffusion, which has been analyzed thoroughly in multiplex networks\cite{gomez2013diffusion,sole2013spectral} %(i.e., it networks of $\mathbb{S}\mathbb{N}\mathbb{I}$ type), 
reveals an intriguing and unexpected phenomenon: diffusion can be faster in a multiplex network than in any of the layers considered independently\cite{gomez2013diffusion}. 

One can understand diffusion in multiplex networks in terms of the spectral properties of a Laplacian tensor (in particular, we consider the type of Laplacian that is known in graph theory as the ``combinatorial Laplacian'' \cite{chung1997}), obtained from the adjacency tensor of \rev{a} multilayer network, that governs the diffusive dynamics. One first ``flattens''\cite{kolda2009} --- without loss of information, provided one keeps the layer labels --- the Laplacian tensor\cite{dedomenico2013mathematical} into a special lower-order tensor called ``supra-Laplacian matrix''. The supra-Laplacian matrix has a block-diagonal structure, where diagonal blocks encode the associated Laplacian matrices corresponding to each layer separately and off-diagonal blocks encode inter-layer connections. \rev{The supra-Laplacian matrix was initially presented in the literature as a matrix for a multilayer network that includes both intra-layer edges and inter-layer edges\cite{gomez2013diffusion}.}

\begin{figure}
\centering
\includegraphics[width=\textwidth]{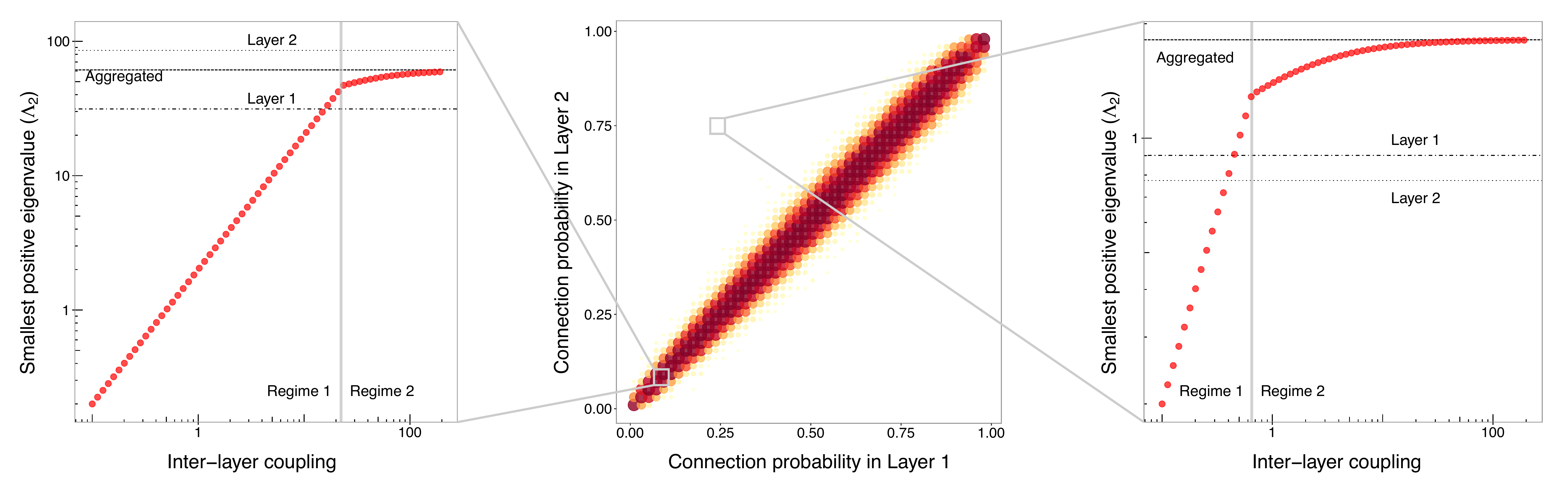}
\caption{\label{fig:diffusion}\textbf{Single dynamics on a multilayer network.} \rev{The speed of diffusion dynamics in a multilayer network is characterized by the second smallest eigenvalue $\Lambda_{2}$ of a Laplacian tensor. We consider a pair of coupled Erd\H{o}s--R\'{e}nyi (ER) networks in which we independently vary the probabilities $p_1,p_2 \in [0,1]$ to connect two nodes within the same layer between. 
The condition\cite{gomez2013diffusion} to observe faster diffusion in the multilayer network than diffusion in each layer separately is $\Lambda_{2}^{\text{multiplex}}\geq \max\{\Lambda_{2}^{\text{layer}~1}, \Lambda_{2}^{\text{layer}~2}\}$. In the central panel, we see that the condition is satisfied when the two layers have similar edge-connection probabilities (i.e., $p_1 \approx p_2)$. We set the inter-layer connections between each node and its replica in the other layer to a weight $\omega \in [0.1,110]$. In the left panel, we show the behavior of $\Lambda_{2}$ as a function of the inter-layer coupling weight $\omega$. A sharp change in the value of the $\Lambda_{2}$ separates two different regimes that correspond to different structural properties of the multilayer network.}
}
\end{figure}

The time scale of diffusion is controlled by \rev{the} smallest positive eigenvalue $\Lambda_{2}$ of the supra-Laplacian matrix. In Fig.~\ref{fig:diffusion}, we show a representative result that \rev{conveys} the existence of two distinct regimes in multiplex networks as a function of \rev{the inter-layer coupling strength}. The regimes illustrate how multilayer structure can influence the outcome of a physical process. For small values of the inter-layer coupling, the multilayer structure slows down the diffusion; for large values, the diffusion speed converges to the mean diffusion speed of the superposition of layers. In many cases, the diffusion in the superposition is faster than that in any of the separate layers. \rev{These findings are a direct consequence of the emergence of more paths between every pair of nodes due to the multilayer structure.} 
The transition between the two regimes is a structural transition\cite{radicchi2013abrupt}, a characteristic of multilayer networks that can \rev{also arise} in other contexts\cite{bazzi2015,taylor2015}.

The above phenomenology can also occur in discrete processes. Perhaps the most canonical examples of discrete dynamics are random walks, which are used to model Markovian dynamics on monolayer networks and which have yielded numerous insights over the last several decades\cite{aldous2002,Gleich:2014tn}. In a random walk, a discretized form of diffusion, a walker jumps between nodes through available connections. In a multilayer network, the available connections include layer switching via an inter-layer edge, a transition that has no counterpart in monolayer networks and which enriches random-walk dynamics\cite{mucha2010community,radicchi2014driving,dedomenico2014navigability}. An important physical insight of the interplay between multilayer structure and the dynamics of random walkers is ``navigability''\cite{dedomenico2014navigability}, which we take to be the mean fraction of nodes that are visited by a random walker in a finite time, which (similar to the case of continuous diffusion) can be larger than the navigability of an aggregated network of layers. In terms of navigability, multilayer networks are more resilient to uniformly random failures than their individual layers, and such resilience arises directly from the interplay between the multilayer structure and the dynamical process.

Another physical phenomenon that arises in multilayer networks is related to congestion, which arises from a balance between flow over network structures and the capacity of such structures to support flow. Congestion in networks was analyzed many years ago in the physics literature \cite{guimera2002optimal,zhao2005onset,echenique2005dynamics}, \rev{but it has been studied only recently} in multilayer networks \cite{tan2014traffic,sole2016PRL}, which can be used to model multimodal transportation systems. It is now known that the multilayer structure of a multiplex network can induce congestion even when a system would remain decongested in each layer independently\cite{sole2016PRL}. 

\rev{\header{Coupled dynamics}}

\rev{Coupled} dynamical processes are a second archetypical family of dynamics in which multilayer structure plays a crucial role. \rev{Thus far, the most thoroughly studied} examples are coupled spreading processes, which are crucial for understanding phenomena such as the spreading dynamics of two concurrent diseases in \rev{two-layer multiplex} networks\cite{dickison2012epidemics,cozzo2013contact,sanz2014dynamics,salehi2015spreading,ferraz2015epidemic} and spread of disease coupled with the spread of information or behavior\cite{bauch2015,funk2015,Funk2009,granell2013interplay,granell2014competing,lima2015disease}. We illustrate two basic effects: (i) \rev{two} spreading processes can enhance each other (e.g., one disease facilitates infection by the other\cite{sanz2014dynamics}), and (ii) one process can inhibit the spread of the other (e.g., a disease can inhibit infection by another disease\cite{sanz2014dynamics} or the spreading of awareness about a disease can inhibit the spread of the disease\cite{granell2013interplay}). Interacting spreading processes also exhibit other fascinating dynamics, and multilayer networks provide a natural means to explore them \cite{bauch2015}.

The \rev{above} phenomenology is characterized by the existence of a curve of critical points that separate \rev{endemic} and non-endemic phases \rev{of a disease}. This curve exhibits a crossover between two different regimes: (i) a regime in which the critical properties of one spreading process are independent of the other, and (ii) a regime in which the critical properties of one spreading process do depend on those of the other. The point at which this crossover occurs is called a ``metacritical'' point.

\begin{figure}
\centering
\includegraphics[width=\textwidth]{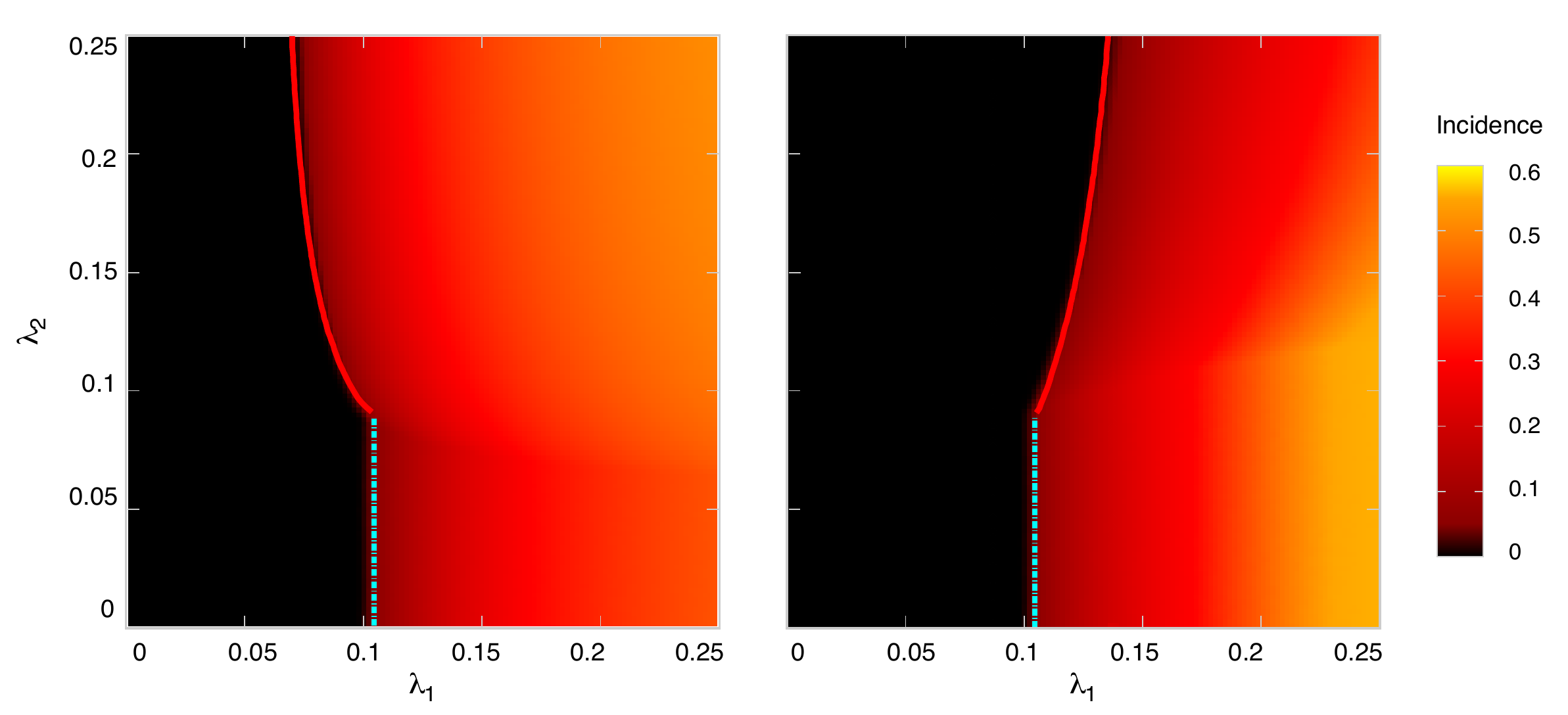}
\caption{\label{fig:spreading}\textbf{\rev{Coupled} dynamics on multilayer networks.} Two (left) reciprocally-enhanced and (right) reciprocally-inhibited disease-spreading processes of susceptible--infected--susceptible (SIS) type. We \rev{compute} these diagrams for multiplex networks formed by two layers of 5000-node Erd\H{o}s--R\'enyi graphs of 5000 with mean intra-layer degree $\langle k \rangle=7$. The colors in the figure represent the prevalence levels of the diseases at a steady state of Monte-Carlo simulations. Note the emergence of a curve of critical points (at a ``metacritical point'') in which the spreading in one layer depends on the spreading in the other. }
\end{figure}

In Fig.~\ref{fig:spreading}, we show (left) a phase diagram of \rev{disease} incidence in one layer of two reciprocally enhanced disease spreading processes; and (right) a phase diagram of the incidence in one layer of \rev{an inhibitory} disease spreading process \rev{affecting another disease.} The metacritical point delineates the transition between independence (dashed line) and dependence (solid curve) of the critical properties of the two processes.

%%%%%%%%%%%%%%%%%%%%%%%%%%%%%%%%%%%%%%%%%%%%%%%%%%%%%%%%%%%%
%%
%%                                       C O N C L U S I O N S    A N D    O U T L O O K  
%%
%%%%%%%%%%%%%%%%%%%%%%%%%%%%%%%%%%%%%%%%%%%%%%%%%%%%%%%%%%%%

\rev{\section*{Conclusions and perspectives}}

In most natural and engineered systems, entities interact with each other in complicated patterns that include multiple types of relationships and/or multiple subsystems, change in time, and incorporate other complications. The theory of multilayer networks seeks to take such features into account to improve our understanding of such complex systems. 

In the last few years, there have been intense efforts to generalize traditional network theory by developing and validating a framework to study multilayer systems in a comprehensive fashion. 
The implications of multilayer network structure and dynamics are now being explored in fields as diverse as neuroscience\cite{betzel2016,braun2015reconfiguration,dedomenico2016brain}, transportation\cite{gallotti2014anatomy,gallotti2015}, ecology\cite{pilosof2015}, granular materials\cite{lia2016}, evolutionary game theory \cite{wang2015evolutionary}, and many others.  For instance, in ecological networks, different layers might encode different types of interaction --- e.g.~ trophic and non-trophic --- or different spatial patches (or different temporal snapshots), where the same interaction may or may not appear \cite{pilosof2015}. In human brain networks, different layers might encode functional connectivity corresponding to specific frequency bands, with inter-layer connections encoding cross-frequency interactions \cite{dedomenico2016brain}. In gene interaction networks, layers might correspond to different genetic interactions (e.g.,~suppressive, additive, or based on physical or chemical associations)\cite{dedomenico2015structural,latora2015biogrid}. In financial networks, layers might represent different interdependent networks of entities\cite{bargigli2015multiplex} --- e.g.,~banking networks and commercial firms --- or different trade relationships among legal entities, ranging from individuals to countries. Despite considerable progress in the last few years\cite{kivela2014multilayer,boccaletti2014structure}, much remains to be done to obtain a deep understanding of the new physics of multilayer network structure and multilayer network dynamics (both dynamics of and dynamics on such networks). In seeking such a deep understanding, it is crucial to underscore the inextricable interdependence of the structure and dynamics of networks.

Recent efforts have revealed fundamental new physics in multilayer networks. The richer types of spreading and random-walk dynamics can lead to enhanced navigability, induced congestion, and the emergence of new critical properties. 
Such new phenomena also have a major impact on practical goals such as coarse-graining networks to examine mesoscale features and evaluating the importance of nodes --- two goals that \rev{date to the beginning of investigations of networks\cite{faust,Newman2010Book,kochen}. For multilayer networks to achieve their vast potential, there remain crucial problems to address. For example, from a structural point of view, it is much easier to measure edge weights reliably for intra-layer edges than for inter-layer edges. Moreover, inter-layer edges not only play a different role from intra-layer ones, but they also play different roles in different applications, and the research community is only scratching the surface of the implications of their presence and the new phenomena to which they lead. For example, how to infer or impose inter-layer edges (and their associated meaning) is a major challenge in many applications of social networks, where an inter-layer connection could exploit the fact of changing from a social platform to another in time as the probability of switching. This can be even more complicated in many types of biological networks (e.g., when considering protein and genetic interactions). We know that different layers are not independent of each other, but it is much more difficult to quantify and measure the weights of the dependencies in a meaningful way. Another major challenge is to understand the propagation of dynamical correlations, due to network structure, across different layers, which affects not only spreading processes but dynamical systems more generally.}
 
\rev{Although our manuscript only addresses physical phenomena related to spreading processes, other dynamical processes also pose extremely fascinating questions\cite{porter2016}. One important example is synchronization, although there are many others (e.g., opinion models, games, and more). A few studies with particular setups have made good progress on multilayer synchronization (see, e.g., \cite{Sorrentino12,Tang14,xiyun15,Sevilla15}), but there phenomenology is very rich, and it will require the development of solid theoretical grounding to study synchronization manifolds, stability analysis, transient dynamics, and more. Additionally, one can build on diffusion dynamics to study reaction--diffusion systems in multilayer networks \cite{turing2014,turing2015}.}

\rev{A particularly promising approach in network theory that will have a major impact on future studies of multilayer networks is the analysis of network structure that arises from latent geometrical spaces\cite{Marian10,Brockmann13,wu15}. Observed connectivity in networks often depends on space\cite{Barthelemy2011} --- either through explicit constraints or by influencing the existence probability and weights of edges --- and thus on the distance in that space. Either or both of the latent space (e.g., people with connections on more layers can lead to a higher probability of observing an edge between them\cite{kempe-multiplex}) and an associated observed network connections can have a multilayer structure. Such explicit use of geometry also allows the possibility of incorporating more continuum types of analyses to accompany the traditional discrete approaches to studying networks. We thus assert that techniques from both geometry and statistics will be crucial for scrutinizing dynamical processes on multilayer networks.}

\rev{The study of multilayer networks is in its infancy, and new emergent physical phenomena that arise from the interaction of such networks and the dynamical processes on top of them are waiting to be discovered.}

%%%%%%%%%

\section*{Contributions}

All of the authors wrote the paper and contributed equally to the production of the manuscript.

%%%%%

\section*{Competing financial interests}

The authors declare no competing financial interests.

\section*{Acknowledgements}

\noindent 
All authors were funded by FET-Proactive project PLEXMATH (FP7-ICT-2011-8; grant \# 317614) funded by the European Commission. MDD acknowledges financial support from the Spanish program Juan de la Cierva (IJCI-2014-20225). CG acknowledges financial support from a James~S.\ McDonnell Foundation postdoctoral fellowship. AA  acknowledges financial support from the ICREA Academia, the James~S.\ McDonnell Foundation, and FIS2015-38266. MAP acknowledges a grant (EP/J001759/1) from the EPSRC. The authors acknowledge help from Serafina Agnello on the creative design of figures. 

%%%%

\bibliographystyle{naturemag}
%\bibliography{multilayer_progress6}

\begin{figure}
\centering
\includegraphics[width=\textwidth]{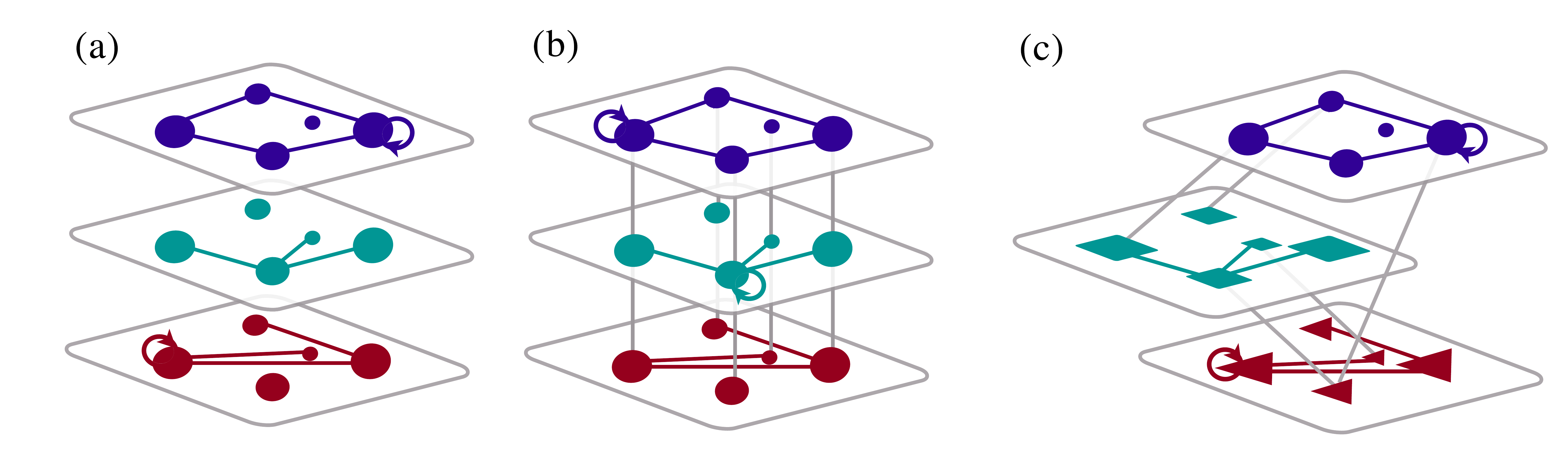}
\caption{\label{fig:multilayer_types}{\bf Multilayer networks}. \rev{({\bf a}) An edge-colored multigraph, in which nodes can be connected by different types (i.e., colors) of interactions. In this example, there are no inter-layer edges. ({\bf b}) A multiplex network, which consists of an edge-colored multigraph along with inter-layer edges that connect entities with their replicas on other layers. ({\bf c}) An interdependent network, in which each layer contains nodes of a different type (circles, squares, and triangles) and includes inter-layer edges to nodes in other layers; in this case, inter-layer edges can occur either between entities and their replicas or between different entities.}
}
\end{figure}

\end{document}